\documentclass[aps,prl,superscriptaddress,amsmath,amssymb,amsfonts,twocolumn,showpacs]{revtex4}

\usepackage{graphicx}
\usepackage{bm}
\usepackage{bbold}

\newcommand{\bea}{\begin{eqnarray}}

\newcommand{\eea}{\end{eqnarray}}

\newcommand{\beq}{\begin{equation}}

\newcommand{\eeq}{\end{equation}}

\newlength{\textwidthm}

\begin{document}

\def \tr{{\mbox{tr~}}}
\def \ra{{\rightarrow}}
\def \ua{{\uparrow}}
\def \da{{\downarrow}}
\def \ba{\begin{array}}
\def \ea{\end{array}}
\def \nn{\nonumber}
\def \l{\left}
\def \rg{\right}
\def \half{{\frac{1}{2}}}
\def \etal{{\it {et al}}}
\def \cH{{\cal{H}}}
\def \cE{{\cal{E}}}
\def \cK{{\cal{K}}}
\def \cM{{\cal{M}}}
\def \cN{{\cal{N}}}
\def \cQ{{\cal Q}}
\def \cI{{\cal I}}
\def \cV{{\cal V}}
\def \cG{{\cal G}}
\def \cF{{\cal F}}
\def \cZ{{\cal Z}}
\def \cC{{\cal C}}
\def \cO{{\cal O}}
\def \bS{{\bf S}}
\def \bI{{\bf I}}
\def \bL{{\bf L}}
\def \bG{{\bf G}}
\def \bQ{{\bf Q}}
\def \bK{{\bf K}}
\def \bR{{\bf R}}
\def \br{{\bf r}}
\def \bu{{\bf u}}
\def \bp{{\bf p}}
\def \bq{{\bf q}}
\def \bk{{\bf k}}
\def \bz{{\bf z}}
\def \bx{{\bf x}}
\def \bpsi{{\bar{\psi}}}
\def \tJ{{\tilde{J}}}
\def \tK{{\tilde{\kappa}}}
\def \W{{\Omega}}
\def \lam{{\lambda}}
\def \L{{\Lambda}}
\def \a{{\alpha}}
\def \t{{\theta}}
\def \T{{\Theta}}
\def \b{{\beta}}
\def \g{{\gamma}}
\def \D{{\Delta}}
\def \d{{\delta}}
\def \w{{\omega}}
\def \s{{\sigma}}
\def \f{{\phi}}
\def \F{{\Phi}}
\def \x{{\chi}}
\def \e{{\epsilon}}
\def \h{{\eta}}
\def \G{{\Gamma}}
\def \z{{\zeta}}
\def \hatt{{\hat{\t}}}
\def \hn{{\bar{n}}}
\def \vk{{{\bf k}}}
\def \vq{{{\bf q}}}
\def \vp{{{\bf p}}}
\def \gk{{\g_{\vk}}}
\def \tp{{\tilde{t}}}
\def \nd{{^{\vphantom{\dagger}}}}
\def \yd{^\dagger}

\title{Electronic compressibility of a graphene bilayer}

\author{S.~Viola~Kusminskiy, Johan~Nilsson, D.~K.~Campbell, and A.~H.~Castro~Neto}

\affiliation{Department of Physics, Boston University, 590 Commonwealth Ave., Boston, MA 02215}

%\date{\today}
 
\begin{abstract}
We calculate the electronic compressibility arising from electron-electron
interactions for a graphene bilayer within the Hartree-Fock approximation. We
show
that, due to the chiral nature
of the particles in this system, the compressibility is rather 
different from those of either the
two-dimensional electron gas or ordinary semiconductors. We find that an
inherent competition between the contributions coming from intra-band
exchange interactions (dominant at low densities) and inter-band interactions (dominant 
at moderate densities) leads to a non-monotonic behavior
of the compressibility as a function of carrier density. 
\end{abstract}


\maketitle

The recently developed experimental capability of isolating and manipulating
an arbitrary number of graphene layers \cite{Geim_review} has
attracted considerable attention  both for its impact on basic science \cite{pw} and for
the tantalizing potential technological applications. The graphene
bilayer is particularly interesting because of the possibility of opening - and controlling - a
gap in the electronic spectrum by applying an external electric field
\cite{Castro06,johan_prl,ohta06,McCann06}. This is not possible for the single layer graphene. The 
bilayer, therefore, while inheriting many of the peculiar electronic characteristics of
the monolayer due to its chiral Dirac fermion (though massive) spectrum, 
has the added virtue of being capable of acting
as an electronic switch. It is thus essential to obtain a
comprehensive characterization of this material. While some transport
experiments are available \cite{Novoselov2006_bilayer_short},
thermodynamic measurements are largely lacking. Among the thermodynamic
quantities to be measured, the electronic compressibility $\kappa$ stands out
as an excellent tool to provide insight into the many-body interactions
present in this material. $\kappa$ can be obtained from the ground state 
energy as:
\beq
\label{kappadef} \kappa^{-1} = n_e^2 (\partial^2 E/\partial n_e^2) \, ,
\eeq
where $E$ is the ground state energy per unit area, and 
$n_e$ is the electronic density.
The electronic compressibility of a single layer graphene has been
recently measured \cite{JMartin_short}, and its behavior, besides
being remarkably different from that of the usual two-dimensional gas
(2DEG), seems to indicate that contributions from Coulomb interactions
are either very weak or cancel out. Hartree-Fock
\cite{dasSarmaK} and Random Phase Approximation (RPA)
\cite{macdonald} calculations predict a correction between 10$\%$ and
20$\%$ to the free theory for experimentally realized dopings. This correction increases logarithmically as the doping is lowered. It is natural then to ask what role
interactions play in the bilayer. In many aspects, the bilayer
graphene closely resembles the 2DEG, as described below. Hence, the
bilayer system provides an opportunity to isolate the effects arising
from its single layer constituents, from those occurring in an
ordinary 2DEG.  In particular, the issue of the chirality, which is so
important for weak-localization physics \cite{ando}, is the main
difference between these two systems, and as we will show, plays an
important role in the many-body physics of the bilayer.  For small
doping, the bilayer can be mapped approximately to a chiral
two-dimensional massive fermionic system with parabolic bands
\cite{McCannFalko06_short,Nilsson06a_short}, 
in contrast with the
massless, cone-like dispersion found in the monolayer. This limit is
useful to compare the behavior of the bilayer (with its chirality) to
that of the ordinary 2DEG, where experiments have shown that
interactions play a dominant role, making the proper compressibility
negative for small electron densities \cite{Eisenstein94} as opposed
to a positive constant given by the non-interacting model. This
behavior is already present at the Hartree-Fock level. Due to the
aforementioned mapping, {\it a priori} it is reasonable to expect that
the effect of electron-electron interactions would be observable in
the graphene bilayer.

In this paper we calculate within the Hartree-Fock approximation the dependence of the inverse
compressibility on the Fermi vector $k_F$ using both a full, four-band (4B) model and 
the two-band (2B) approximation, which is valid for very small doping. We show that the most
important qualitative signatures of the compressibility are already present
at the 2B model level but that the 4B calculation,  while more cumbersome, reveals finer features.

Throughout this paper we will refer (loosely) to the quantity $\tK^{-1}=\partial \mu/\partial n_e$ as the
``inverse compressibility''. Here $\mu$ stands
for the chemical potential of the system. $\tK$ differs by a factor of $n_e^2$
from $\kappa$ in (\ref{kappadef}). This is appropriate since $\tK^{-1}$ is
usually the actual experimentally measured quantity. The density of electrons is given by
$n_e=g_Sg_v k_F^2/(4\pi)$, with $g_s=2$, $g_v=2$ being the spin
and valley degeneracy, respectively. 
In the following we will consider the case of small doping
but outside the range of ferromagnetic instability that is found at extremely
low doping \cite {Nilsson06a_short}. At the Hartree-Fock
level, the ground state energy is given by $E=K+E_{ex}$, where $K$ stands for
the kinetic energy and $E_{ex}$ is the exchange energy per unit of area.

\begin{figure}
  \centering
  \includegraphics[width=7cm,height=4cm]{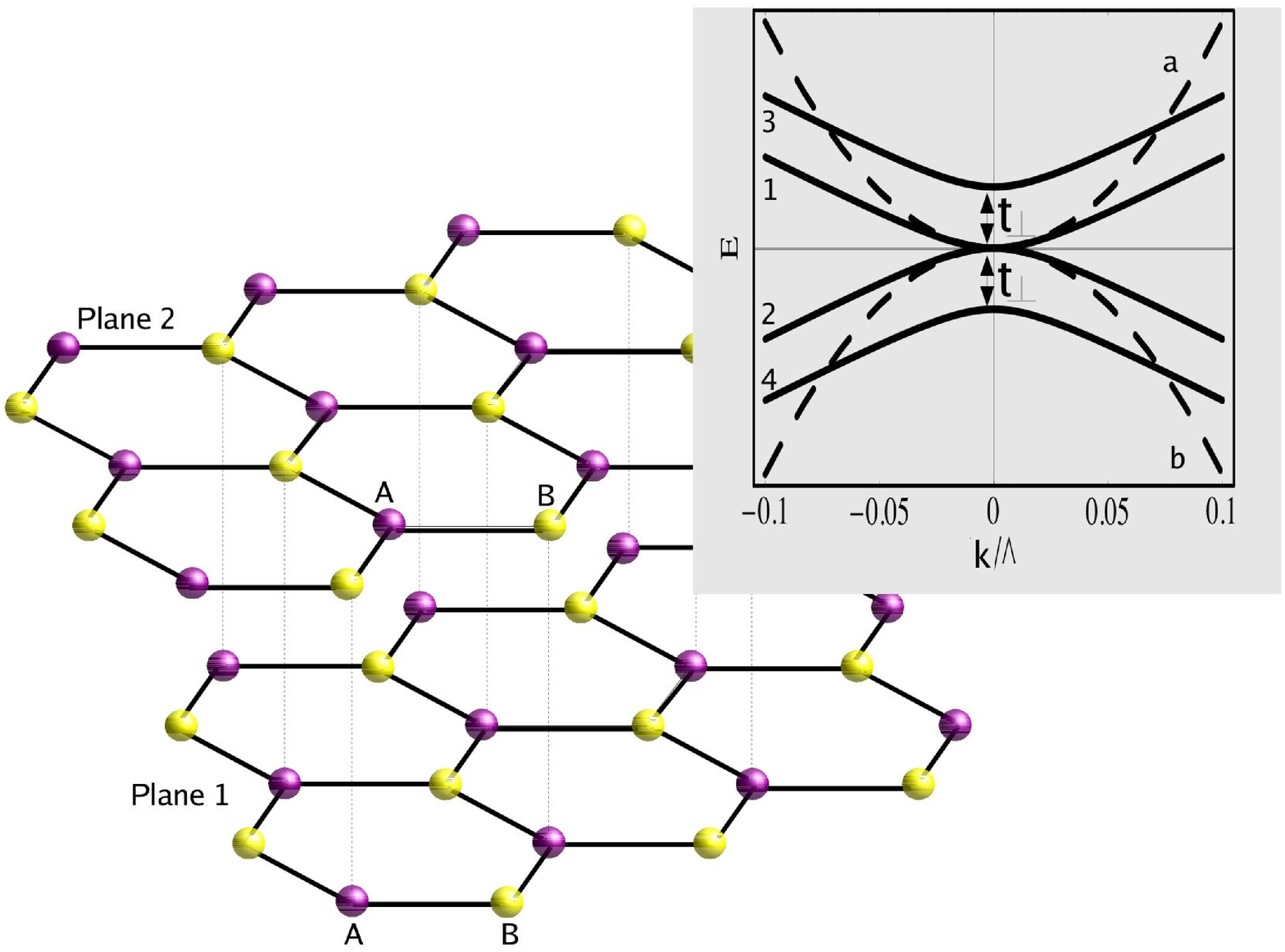}
  \caption{(Color online) Graphene bilayer lattice showing the two underlying sublattices
    A (purple) and B (yellow). \textit{Inset}: 4B dispersion (continuous line) and 2B approximation (dashed) 
as a function of momentum $k$. Momentum is in units of the cutoff $\L$, measured from the K point.}
  \label{fig:latt}
\end{figure}

A graphene bilayer consists of two planes of graphene stacked as shown in
Fig. (\ref{fig:latt}). The kinetic term of the Hamiltonian can be written, in the
nearest neighbor tight binding
approximation \cite{Wallace47} by expanding around the K, K' points of the
Brillouin zone, as $\cH_{kin}=\sum_\cQ \psi_\cQ \yd \cK(\vp)\psi_\cQ$ with $\psi_\cQ
\yd=(c\yd_{\vp,A_1,\s,a},c\yd_{\vp,B_1,\s,a},c\yd_{\vp,B_2,\s,a},c\yd_{\vp,A_2,\s,a})$
where $a$ labels the valley, $\s$ the spin and $A_i$, $B_i$ denotes the
sublattice in the plane $i=1, 2$. $\sum_\cQ$ represents the sum over all the
indices. The kinetic energy matrix is given by (we use units such that $\hbar =1$):
\beq\label{kinmatrix}
\cK(\vp)=\l(\ba{cccc}
0 & pe^{i\f(\vp)}& -t_\perp &0\\
pe^{-i\f(\vp)}&0 & 0 & 0\\
-t_\perp &0 & 0 & pe^{-i\f(\vp)}\\
0 & 0& pe^{i\f(\vp)}& 0
\ea\rg) \, ,
\eeq
where $\tan{\f(\vp)}=p_y/p_x$, $t_{\perp}=0.35 \,{\rm eV}$ is the inter-layer hopping energy
and we have set 
$v_F =3 t a/2$ ($\approx 6.6 \,{\rm eV}$~\AA) to unity 
($t$ is the intra-layer hopping energy and $a$ the in-plane
carbon-carbon distance). The interaction is given by the 2D Fourier transform
of the 3D Coulomb potential, which is $V_{ip}(\bk)=\frac{2\pi e^2}{\e_o}
\frac{1}{k}$ for the interaction among electrons within the same plane and
$V_{op}(\bk)=\frac{2\pi e^2}{\e_o} \frac{e^{-kd}}{k}$ otherwise, being $d
\approx 3.35$~\AA $\,$ the inter-plane distance.

The kinetic
energy matrix (\ref{kinmatrix}) can be diagonalized by a unitary
transformation $S\yd(\vp)$. The resulting dispersion bands (see
Fig.(\ref{fig:latt})) are: $E_{1}(p)=-\tp+E(p)$,
$E_{2}(p)=\tp-E(p)$, 
$E_{3}(p)=\tp+E(p)$ and $E_{4}(p)=-\tp-E(p)$; being
$E(p)=\sqrt{\tp^2+p^2}$ and $\tp=t_\perp/2$. It is convenient to work with the
symmetric and anti-symmetric combinations of the layer densities,
$\rho_\pm=\rho_1\pm\rho_2$, which can be expressed in the diagonal basis as
$\rho_\a(\vq)=\sum_{\vp}\F\yd(\vp+\vq)\x^\a(\vp+\vq,\vp)\F(\vp)$ with
$\F_(\vp)=S\yd(\vp)\psi(\vp)$ and $\a=\pm$. The $4
\times 4$ matrices $\x^\a$ contain the
information of the overlap due to the change of basis. Then the interaction
Hamiltonian takes the form $\cH_I= 1/(2A) \sum_{\vq\neq0}\sum_\a
\rho_\a(\vq)V_\a(\vq)\rho_\a(\vq)$, and the exchange energy per unit area $A$
can be written in the continuum as: 
\beq \label{Eex1}
\begin{split}\vspace{1mm} E_{ex}=-g_sg_v\half \int
  \frac{d^2\bp}{(2\pi)^2}\frac{d^2\bq}{(2\pi)^2}& \sum_{\a,i,j}
  \x_{ij}^\a(\bq,\bp)\x_{ji}^\a(\bp,\bq)\\ &n_{i}(q)n_{j}(p)V_\a(\bq-\bp) \, ,
\end{split}
\eeq
where $i,j=1,...4$; $n_{i}(q)=<\F_i
\yd(\vq)\F_i(\vq)>$ and $V_{\pm}(\bk)=\half(V_{ip}(\bk)\pm V_{op}(\bk))$. 
The occupation factors are given by
$n_{1}(q)=\T{(k_F-q)}$ [$n_{1}(q)=0$], $n_{2}(q)=1$ [$n_{2}(q)=1-\T{(k_F-q)}$], $n_{3}(q)=0$ and
$n_{4}(q)=1$ in the case of electron [hole] doping. This model however
requires a cutoff $\L$ of the order of the
inverse of the lattice parameter. 

Being simpler to work with, and widely used as a starting point for calculations in the graphene bilayer, we start our analysis with the approximate 2B model that can be constructed at low energies by performing
degenerate perturbation theory \cite{McCannFalko06_short,Nilsson06a_short}. This results in an effective kinetic Hamiltonian: 
\beq \begin{split}\vspace{2mm}
  \cH_{kin} &=\sum_\cQ \frac{p^2}{2\tp}
  \tilde{\psi}_\cQ \yd \l(\ba{cc} 0 &e^{-2i\f(\vp)}
\\
  e^{2i\f(\vp)} & 0\ea\rg)\tilde{\psi}_\cQ \, ,
\end{split} 
\eeq 
with $\tilde{\psi}_\cQ\yd=(c\yd_{\vp,B_1,\s,a},c\yd_{\vp,A_2,\s,a})$. 
The result of the approximation is an effective model with opposite parabolic dispersion
bands of energy $E_{{\rm a}}=p^2/(2\tp)$, $E_{{\rm b}}=-p^2/(2\tp)$ as shown in
Fig.(\ref{fig:latt}). The effective kinetic energy per unit area then is
given by $K=(k_F^4-\L^4)/(4 \pi \tp)$ giving a kinetic
contribution to the inverse compressibility of $\tK^{-1}_K=\pi/(2\tp)$.

In this reduced Hilbert space Eq. (\ref{Eex1}) is still valid, but this time the
$\x^\a(\vp,\vq)$ are $2 \times 2$ matrices. Combining all the contributions and
re-inserting the units, we find the 
total inverse compressibility  $\tK^{-1}$ in the 2B model to be given by the
expression:
\beq \label{totalK}\begin{split}\vspace{1mm}
\tK^{-1}&=\frac{
v_F\hbar}{\L}\l[\frac{\pi}{2\tp}-\frac{g}{4k_F}\int_0^{\pi}d\t\int_0^1pdp
\frac{\partial}{\partial k_F}\rg.\\\vspace{1mm}&\l[
\frac{k_F}{r(p,1,\t)}\l(1+\cos{2\t}e^{-dk_Fr(p,1,\t)}\rg)\rg. \\ &
+\l.\l. \frac{1}{r(p,k_F,\t)}\l(\pm 1-\cos{2\t}e^{-dr(p,k_F,\t)}\rg)\rg]\rg] \, 
\end{split}
\eeq
Here we have defined $r(p,q,\t)=\sqrt{p^2+q^2-2pq\cos{\t}}$, $\tp$, $k_F$
and $1/d$ are in units of $\L$, and $g= e^2/(v_F\e_0)$ is the graphene
coupling strength. The $\pm$ indicates the
expression for electrons ($+$) or holes ($-$). The differing term however is roughly a constant and can be neglected for small
doping. Therefore, in what follows we will use the results for electron doping.

\begin{figure}
 \centering
 \includegraphics[width=6cm,height=8cm,angle=-90]{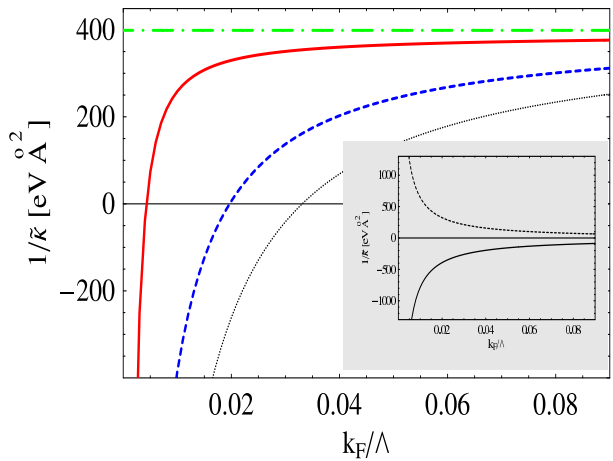}
 \caption{(Color online) Inverse compressibility vs. Fermi wave vector in the 2-band
   approximation. Total: solid line (red), intra-band plus
   kinetic contribution: dashed (blue), kinetic: dash/dot (green), 2DEG:
   dotted (black). The inset depicts the contribution of intra-band exchange
   (solid line) and inter-band exchange (dashed). }
  \label{fig:figg1}
 \end{figure}

Fig.(\ref{fig:figg1}) shows a plot of (\ref{totalK}) as a function of
$k_F$ for $\tp/\L=0.026$, and $d\L=3.7$ ($\L=1.06 \approx
1$~\AA$^{-1}$). As can be seen, for very small doping the
compressibility changes sign, becoming negative and divergent. This
behavior, as mentioned previously, is also observed in the 2DEG. It is
instructive to discriminate between inter and intra-band
contributions. Fig.(\ref{fig:figg1}) also depicts the inverse
compressibility when only the intra-band transitions are considered
(as well as the kinetic term, of course). From the difference with the
curve for the total inverse compressibility, we can conclude that the
inter-band contribution tends to move the negative region to smaller
densities. The overall effect of the inter-band transitions then is to
enhance $\tK^{-1}$, therefore reducing the compressibility.  This can
be seen clearly from the inset in Fig.(\ref{fig:figg1}), where the
contribution from intra-band transitions is negative while that from
inter-band transitions is positive. Apart from the sign, both present
a similar behavior and are comparable in magnitude.  In the 2B
approximation, the kinetic contribution, as in the 2DEG case, is
independent of the electronic density and is also plotted in
Fig.(\ref{fig:figg1}) for reference. Therefore, the resulting total
compressibility will be given by a competition of the two
contributions. 
This difference in sign between inter and intra-band
contributions is analogous to the one present in monolayer graphene
\cite{JMartin_short} and, whitin the 2B approximation, it can be
interpreted in terms of the chirality of the quasiparticles. Intra
(inter)-band exchange corresponds to interactions between particles of
the same (opposite) chirality. Remarkably though, while for the monolayer the total exchange contribution is positive, for the bilayer is negative.

We can also compare with the usual result for the 2DEG. 
For this we start from the expression for the chemical potential
(see for example \cite{vignale}) $\mu= \partial[n\epsilon(n)]/\partial n$, 
where $n= k_F^2/\pi$ is the electronic density (taking into
account spin and valley degeneracy) and $\epsilon$ is the ground state
energy per electron. If we consider only kinetic and exchange energy,
$\epsilon=\epsilon_0+\epsilon_{ex}$ and for the 2DEG
$\epsilon_0= k_F^2/(4 m)$, $ \epsilon_{ex}=- 4 e^2 k_F/(3 \pi)$. Therefore 
$\tK^{-1}_{2DEG}=(\pi/m-2e^2/k_F)/2$. To compare with our case, we identify $\tp\equiv m$ and write
$\tK^{-1}_{2DEG}=v_F/(2\L) \left(\pi/\tp-2g/k_F\right)$. The
corresponding plot is also depicted in Fig.(\ref{fig:figg1}). Within the 2B approximation, the bilayer compressibility behaves
qualitatively similar to that of the 2DEG, although, because of the chiral nature of the
bilayer system, the region of negative values is shifted to
smaller values of the Fermi vector and therefore smaller densities.

\begin{figure}
  \centering
  \includegraphics[width=7cm]{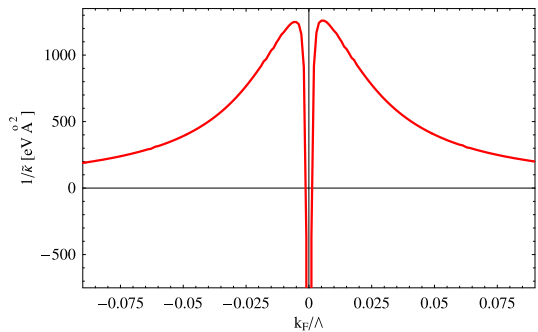}
\caption{(Color online) Inverse compressibility as a function of the Fermi
  wave vector as
  calculated from the 4B model. Negative values of the Fermi vector
  indicate the result for hole doping. }
  \label{fig:figg5}
\end{figure}

As mentioned above, the total compressibility of the bilayer is a product of the competition between inter and intra-band contributions to exchange. However from the inset in Fig.(\ref{fig:figg1}) it transpires that this is a very tight competition and small changes may alter the result. We therefore proceed to analyze the full 4B model. Due to the symmetries of the $\x$ matrices,
expression (\ref{Eex1}) is greatly simplified. The exchange contribution to the compressibility then is given by $$\tK^{-1}_{ex}=-\frac{1}{8\pi k_F}\sum_\a\frac{\partial}{\partial k_F}(D_{++}^\a+D_{+-}^\a)$$ being
\beq \label{DEex1new} \begin{split}\vspace{1.5mm}
D_{++}^\a=&\int_0^{2\pi}d\t\int_0^{k_F}
pdp \, V_\a(|\bk_F-\bp|) |\x_{11}^\a(\bk_F,\bp)|^2 \, ,
\nonumber
\\
\vspace{1.5mm}
D_{+-}^\a=& \int_0^{2\pi}d\t\int_0^\L
pdp \, V_\a(|\bk_F-\bp|)\\ &
\qquad \qquad \quad(|\x_{12}^\a(\bk_F,\bp)|^2+|\x_{14}^\a(\bk_F,\bp)|^2) \, ,
\end{split}
\eeq
where $\t$ is the angle between $\bk_F,\bp$. As the notation suggests, $D_{++}^\a$ corresponds to exchange within
the positive conduction band $1$ while $D_{+-}^\a$ measures the exchange
between the negative filled sea and the conduction band. The calculation for hole
doping is completely analogous, with the overlap elements to be considered for
that case being $|\x_{22}^{\a}(\bk_F,\bp)|^2$ and
$|\x_{24}^{\a}(\bk_F,\bp)|^2$. Since the compressibility involves only
occupied states, its behavior is not symmetric with respect to
particle-hole exchange. Nonetheless, the explicit calculation shows that the
difference is negligible for small doping, being the same as in the 2B case. On the other hand, the kinetic contribution to the inverse compressibility is
independent of the type of carrier and it is easily calculated to be
$\tK^{-1}_K=\pi/[2E(p)]$. The final results for the four band
calculation, by summing all the contributions, is shown in Fig.
(\ref{fig:figg5}).

\begin{figure}
 \centering
 \includegraphics[width=7cm]{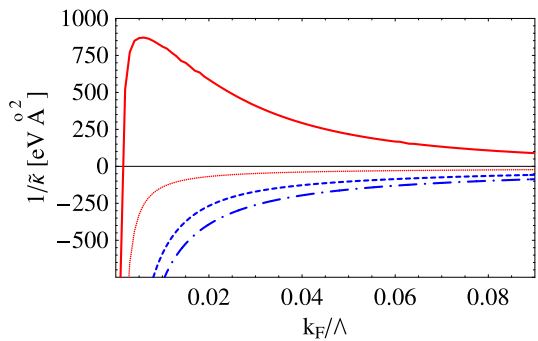}
 \caption{(Color online) Comparison between the exchange contribution for the
   2B and 4B models. Total exchange (red): 4B solid line, 2B
   dotted; intra-band only (blue): 4B dashed, 2B dash/dot.
    }
  \label{fig:figg4}
 \end{figure}

We see that the full model confirms the major
qualitative features found within the 2B approximation.  The
compressibility is negative for small electronic density, diverging in
the limit of $k_F \rightarrow 0$, and the inter-band exchange
contributes to the incompressibility of the system (see Fig. (\ref{fig:figg4})). The difference
observed at larger values of the Fermi momentum is simply due to the
difference in the kinetic term, since it is a constant in the two band
case while in the 4B model it is $\sim 1/k_F$ for large $k_F$.
However, a more detailed comparison reveals a peak in the inverse
compressibility that is not captured in the 2B model.  Overall, the 4B
calculation predicts a more incompressible system in the range for
which $\tK>0$, up to approximately three times larger than the
prediction of the 2B model at the position of the peak. As seen from
Fig. (\ref{fig:figg4}), the bulk of the difference between the 2B
result and the 4B one comes from the inter-band exchange. 

The 4B model
seems to predict a behavior that is a hybrid between the one of a
2DEG, where the total contribution to the compressibility from
exchange is negative, and the graphene monolayer, where it is
positive. 

In conclusion, we have studied the electronic compressibility of a graphene
bilayer within the Hartree-Fock approximation and have found a behavior 
that is remarkably different from the two-dimensional electron gas due to
the presence of inter-band transitions, and also from graphene monolayer. We have shown
that the inverse compressibility is not a monotonic function of the
electronic density and that the effective 2B model gives a
good description of the problem only at very low densities. At intermediate
densities, the four bands are important to explain the behavior of the
compressibility. The non-monotonic behavior of the compressibility obtained with the 4B model is highly unusual. Generally, nonmonotonicity is associated with some external factor, such as confinement or applied magnetic fields. However here it is due solely to intrinsic electronic interactions. The implications of this remain to be understood. On the other hand, the negativity of the compressibility is understood once it is realized that the one involved is not the total compressibility but only that of the electronic gas. The total compressibility will comprise also the positive ionic background, which stabilizes the system. Nonetheless, the negative {\it divergence} of the inverse compressibility,
present in both models for low enough electronic densities, could signal the eventual
onset of Wigner crystallization \cite{dahal07}. These results, as in the case of the single layer graphene
\cite{JMartin_short} can be studied via single electron transistor (SET)
measurements. Our results indicate that the compressibility turns negative at density values of approximately $n_e \approx 10^{11}$/cm$^2$, which borders the current available precision \cite{JMartin_short}. However, being a Hartree-Fock calculation, this can act just as a very rough estimate. We have also neglected the trigonal warping term, which might be of importance at very low densities \cite{McCannFalko06_short}.

We thank G. Giuliani, V. Kotov, B. Uchoa, G. Vignale, and A. Yacoby for 
illuminating discussions. S. V. K. would like to thank M. Reiris for his
invaluable mathematical insight. A. H. C. N. was supported through NSF grant
DMR-0343790.

\end{document}